# Preprint notes

## Title of the article:

AI Ethics in Industry: A Research Framework

## Authors:

Ville Vakkuri, Kai-Kristian Kemell, Pekka Abrahamsson

## Notes:

-This is the author's version of the work

-This is a pre-print of an article accepted to Third Annual Seminar on Technology Ethics

https://future-ethics.utu.fi/iii-seminar-on-technology-ethics/

-The final authenticated version will be link here when published

# AI Ethics in Industry: A Research Framework

Ville Vakkuri[0000-0002-1550-1110], Kai-Kristian Kemell[0000-0002-0225-4560] and Pekka Abrahamsson [0000-0002-4360-2226]

University of Jyväskylä, PO Box 35, FI-40014 Jyväskylä, Finland

ville.vakkuri|kai-kristian.o.kemell|pekka.abrahamsson@jyu.fi

**Abstract.** Artificial Intelligence (AI) systems exert a growing influence on our society. As they become more ubiquitous, their potential negative impacts also become evident through various real-world incidents. Following such early incidents, academic and public discussion on AI ethics has highlighted the need for implementing ethics in AI system development. However, little currently exists in the way of frameworks for understanding the practical implementation of AI ethics. In this paper, we discuss a research framework for implementing AI ethics in industrial settings. The framework presents a starting point for empirical studies into AI ethics but is still being developed further based on its practical utilization.

## 1     Introduction

Artificial Intelligence (AI) and Autonomous Systems (AS) have become increasingly prevalent in software development endeavors, changing the role of ethics in software development. One key difference between conventional software systems and AI systems is that the idea of active users in the context of AI systems can be questioned. More often than not, individuals are simply objects for AI systems that they either perform actions upon or use for data collection purposes. On the other hand, *users* of AI systems are usually organizations as opposed to individuals. This is problematic in terms of consent, not least because one may not even be aware of being used for data collection purposes by an AI.

To this end, existing studies have argued that developing AI/AS is a multi-disciplinary endeavor rather than a simple software engineering one (Charisi et al. 2017). Developers of these systems should be aware of the ethical issues involved in these systems in order to be able to mitigate their potential negative impacts. While discussion on AI ethics among the academia has been active in the recent years, various public voices have also expressed concern over AI/AS following recent real-world incidents (e.g. in relation to unfair systems (Flores, Bechtel & Lowenkamp 2016)).

However, despite the increasing activity in the area of AI ethics, there is currently a gap between research and practice. Few empirical studies on the topic exist, and the state of practice remains largely unknown. The IEEE Ethically Aligned Design guidelines have suggested that they have not been widely adopted by practitioners. Additionally, in a past study, we have presented preliminary results supporting the notion of a gap in the area (Vakkuri, Kemell, Kultanen, Siponen, & Abrahamsson 2019b). Other past studies have shown that developers are not well-informed on ethics in general (McNamara, Smith & Murphy-Hill 2018). This gap points towards a need for tooling and methods in the area, as well as a need for further empirical studies on the topic.

To provide a starting point for bridging the gap between research and practice in terms of empirical research, we present a framework for AI ethics in practice. The framework is built around extant conceptual research in the area of AI ethics, intended to serve a framework for empirical studies into AI ethics. The framework has been utilized in practice to collect empirical data and based on this utilization we discuss the framework in this paper.

The rest of this paper is organized as follows. In section 2 we discuss the theoretical background of the study by going over existing research in the area. Then, in section 3, we present the research framework discussed in this paper. In section 4 we go over the results of an empirical study in which the framework was utilized. In section 5, we discuss the framework and its implications. Section 6 concludes the paper.

## 2     Background: The Current State of AI Ethics

The academic discussion on AI ethics has thus far largely focused on defining the area through central constructs and principles. Thus far, the focus has been on four main principles for AI ethics: transparency (Dignum, 2017; The IEEE Global Initiative on Ethics of Autonomous and Intelligent Systems 2019; Turilli & Floridi 2009), accountability (Dignum 2017; The IEEE Global Initiative on Ethics of Autonomous and Intelligent Systems 2019), responsibility (Dignum 2017), and fairness (e.g. (Flores et al. 2016)). However, not all four of these values are universally agreed to form the core of AI ethics (e.g. (Morley, Floridi, Kinsey & Elhalal 2019)) and effectiveness of using values or principles to approach AI Ethics has been criticized in and of itself (Mittelstadt 2019).

Various real-world incidents out on the field (e.g., (Reuters 2019)) have recently began to spark public discussion on AI ethics. This has led to governments, standardization institutions, and practitioner organizations reacting by producing their own demands and guidelines for involving ethics into AI development, with many guidelines and regulations in the works.  Countries such as France (Villani et al., 2018), Germany (*Ethics commission's complete report on automated and connected driving* 2017) and Finland (*Finland's age of artificial intelligence report* 2017) have emphasized the role of ethics in AI /AS. On an international level, the EU began to draft its own AI ethics guidelines which were presented in April 2019 (AI HLEG 2019). Moreover, the IEEE P7000™ Standards Working Groups ISO has founded its own standardization subcommittee (ISO/IEC JTC 1/SC 42 artificial intelligence.). Finally, larger practitioner organizations have also presented their own guidelines concerning ethics in AI (e.g., Google guidelines (Pichai 2018)), Intel's recommendations for public policy principles on AI (Rao 2017), Microsoft's guidelines for conversational bots (Microsoft, 2018)).

Attempts to bring this on-going academic discussion out on the field have been primarily made in the form of guidelines and principles lacking practices to implement them (Morley et al. 2019). Out of these guidelines, perhaps the most prominent ones up until now have been the IEEE guidelines for Ethically Aligned Design (The IEEE Global Initiative on Ethics of Autonomous and Intelligent Systems 2019), born from the IEEE Global Initiative on Ethics of Autonomous and Intelligent Systems alongside its IEEE P7000™ Standards Working Groups, which were branded under the concept of EAD (The IEEE Global Initiative on Ethics of Autonomous and Intelligent Systems, 2019).

These guidelines, however, are unlikely to see large-scale industry adoption based on what we already know about ethical guidelines in IT. In their study on the effects of the ACM ethical guidelines (Gotterbarn et al. 2018), McNamara et al. (2018) discovered that the guidelines had had little impact on developer behavior. The IEEE EAD (The IEEE Global Initiative on Ethics of Autonomous and Intelligent Systems 2019) guidelines already suggest that this is likely to be the case in AI ethics as well, although there currently exists no empirical data to confirm this assumption.

## 3    Research Model

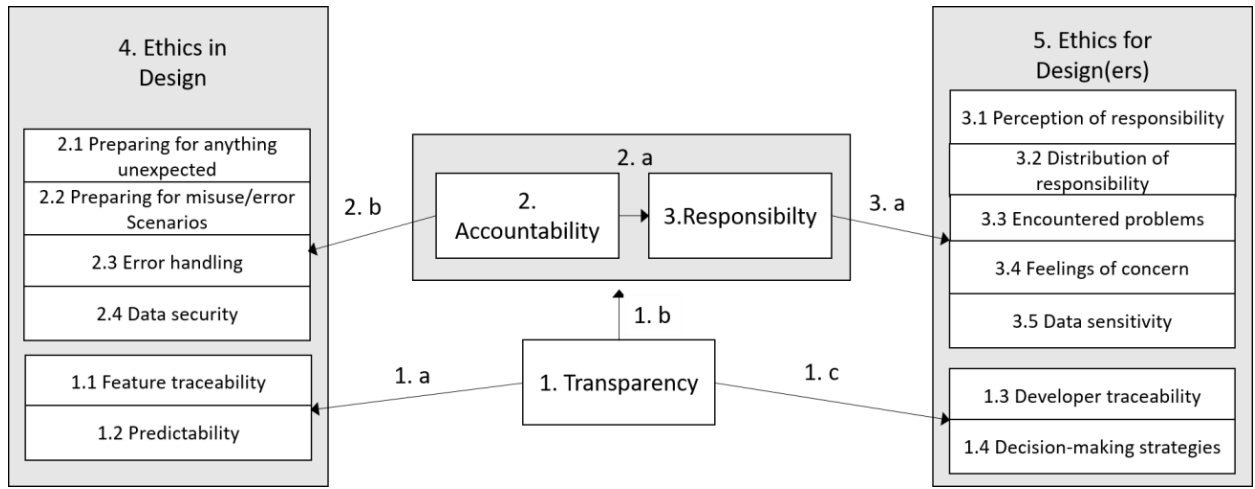

*Figure 1 Research framework*

Academic literature has discussed various principles as a way to address ethics as a part of the development of AI and AI-based systems. Currently, four constructs are considered central ones in AI ethics: Transparency (Dignum, 2017; The IEEE Global Initiative on Ethics of Autonomous and Intelligent Systems., 2019; Turilli & Floridi, 2009), Accountability (Dignum, 2017; Turilli & Floridi, 2009), Responsibility (Dignum, 2017), and Fairness e.g. (Flores et al., 2016). Perhaps notably, a recent EU report (High-Level Expert Group on Artificial Intelligence, 2019) also discussed Trustworthiness as its key construct, a value all systems should aim for, according to the report. Morley et al. (2019) presented an entirely new set of more abstract constructs intended to summarize the existing discussion and the plethora of principles discussed so far in addition to the ones mentioned here. They presented five constructs in the form of: Beneficence, Non-maleficence, Autonomy, Justice, and Explicability.

To categorize the field of AI ethics three categories have been presented: (1)  Ethics by Design (integrating ethics into system behavior); (2) Ethics in Design (software development methods etc. supporting implementation of ethics); and (3) Ethics for Design (standards etc. that ensure the integrity of developers and users) (Dignum, 2018). In this model, we focus on the latter two categories.

Out of the aforementioned four principles that have been proposed to form the basis of ethical development of AI systems, we consider accountability, responsibility, and transparency (the so-called ART principles (Dignum (2017)) a starting point for understanding the involvement of ethics in ICT projects. These three constructs form the basis of ethical AI and attempts to identify their possible relations, as well as relations of other constructs that may be involved in the process.

To make these principles tangible, a subset of constructs in the form of actions (Fig. 1 (1.1-3.5)), discussed in detail in subsection 3.1, was formed under each key concept. These actions were outlined based on the IEEE guidelines for EAD (The IEEE Global Initiative on Ethics of Autonomous and Intelligent Systems, 2019). The actions were split into two categories, Ethics in Design and Ethics for Design(ers), based on Dignum's (2018) typology of AI ethics.

### 3.1 The ART Model

*Transparency* is a key ethical construct that is related to *understanding* AI systems. Dignum (2017) discusses transparency as transparency of AI systems and specifically algorithms and data used. Arguably, transparency is a pro-ethical circumstance that makes it possible to implement AI ethics in the first place (Turilli and Floridi 2009). Without understanding how the system works, it is impossible to understand why it malfunctioned and consequently to establish who is responsible. Additionally, both the EU AI Ethics guidelines (AI HLEG 2019) and EAD guidelines (The IEEE Global Initiative on Ethics of Autonomous and Intelligent Systems 2019) consider transparency an important ethical principle.

In the research framework presented in this paper, we consider transparency not only in relation to AI systems but also in relation to AI systems development. I.e., we also consider it important that we understand what decisions were made, by whom, and why during development. Different practices support this type of transparency (e.g. audits and code documentation (Vakkuri, Kemell, & Abrahamsson 2019a)).

For the system to be considered transparent (line 1.a), feature traceability (1.1) (EAD Principle 5) should be present, and the system should be predictable in its behavior (1.2) (EAD Principles 5 and 6). For development to be considered transparent (line 1.c), the decision-making strategies of the endeavor should be clear (1.4) (EAD Principles 5 and 6), and decisions should be traceable back to individual developers (1.3) (EAD Principles 1, 5, and 6). As a pro-ethical circumstance, transparency also produces the possibility to assess accountability and responsibility (line 1.b) in relation to both development and the system.

*Accountability* refers to determining who is accountable or liable for the decisions made by the AI. Dignum (2017) defines accountability to be the explanation and justification of one's decisions and one's actions to the relevant stakeholders. In the context of this research framework, accountability is used not only in the context of systems, but also in a more general sense. We consider, for example, how various accountability issues (legal, social) were taken into consideration during the development.

As mentioned earlier, transparency is the pro-ethical condition here that makes accountability possible (denoted by line 1.b). We must understand how the system works in order to establish accountability. Similarly, we should be able to determine *why* it works that way by understanding what decisions made during development led to the system working that way. We consider accountability in a broad sense, thus including also legal and social concerns related to the system. Much like transparency, accountability is also considered a key construct in AI ethics (The IEEE Global Initiative on Ethics of Autonomous and Intelligent Systems 2019) and it holds an important role in preventing misuse of AI systems and supporting wellbeing through AI systems.

In our research model, accountability is perceived through the concrete actions of the developers concerning the systems itself, 2.1 Preparing for anything unexpected: (actions that are taken to prevent or control unexpected situation) (EAD Principle 8), 2.2 Preparing for misuse/error scenarios (actions that are taken to prevent or control misuse/error scenarios) (EAD Principles 7 and 8), 2.3 Error handling (practices to deal with errors in software) (EAD Principles 4 and 7) and 2.4 data security (actions taken to ensure cyber security of system and secure handling of data) (EAD Principle 3).

Finally, Dignum (2017) considers *responsibility* a chain of responsibility that links the actions of the system to all the decisions made by its stakeholders. We do not consider this definition to be actionable and instead draw from the EAD guidelines (The IEEE Global Initiative on Ethics of Autonomous and Intelligent Systems 2019) to consider responsibility as an attitude or moral obligation to act ethically. It is thus internally motivated rather than the externally motivated accountability (e.g. legal responsibility).

While accountability relates to the connection between one's decisions and the stakeholders of the system, responsibility is more focused on the internal processes of the developers not necessarily directly related to any one action. In order to act responsibly, one needs to understand the meanings of their actions. Therefore, in the research framework responsibility is perceived through the actions of the developers concerning, 3.1 perception of responsibility (developers have a sense of responsibility and perception what is responsibility in software development) (EAD Principles 2, 4 and 6); 3.2 distribution of responsibility (who is seen responsible e.g. for any harm caused by the system) (EAD Principle 6); 3.3 encountered problems (how errors and error scenarios are tackled and who is responsible for tackling them) (EAD Principles 7 and 8); 3.4 feelings of concern (developers are concerned about issues related to their software); and 3.5 data sensitivity (developers attitude toward data privacy and data security) (EAD Principles 2 and 3).

### *3.2 Operationalizing the Research Framework*

The commitment net model of Abrahamsson (2002) was utilized to analyze the data gathered using this research framework (Vakkuri et al. 2019a). This was done to have an existing theoretical framework to analyze the data with, and especially one aimed at the context of software development.

From this commitment net model, we focused on *concerns* which were analyzed to understand what ethical issues were of interest to the developers. *Actions* were then studied to understand how these concerns were actually tackled, or whether they were tackled at all. In commitment net model, actions are connected to concerns because when actions are taken, they are always driven from concerns (Abrahamsson, 2002). However, concerns can also exist without any actions taken to address them, although this points to a lack of commitment on the matter.

The dynamic between actions and concerns was considered a tangible way to approach the topic of practices for implementing AI ethics. Actions were directly likened to (software development) practices in this context. On the other hand, concerns were considered to be of interest in understanding e.g. whether the developers perhaps wanted to implement ethics but were unable to do so.

In this fashion, existing theories can be used in conjunction with the framework to either make it more actionable for implementing ethics, or for helping analyze or gather data using the framework.

## 4 Empirical Utilization of the Framework

The framework was utilized successfully in a recent study (Vakkuri et al. 2019a). The empirical portion of the focal paper is summarized briefly in this section in order to demonstrate how to benefit from the framework. However, the focus of this paper is on the research framework itself rather than these empirical results.

*4.1 Study Design*

The research framework was utilized to carry out a multiple case study of three case companies. Each company was a software company developing AI solutions for the healthcare industry. More specifically, the case studies focused on one specific project inside each of the case companies.

*Table 1.* Case Information

| Case | Case Description | Respondent[Reference] |
|---|---|---|
| A | Statistical tool for detecting marginalization | Data analyst [R1] |
| A | Statistical tool for detecting marginalization | Consultant [R2] |
| A | Statistical tool for detecting marginalization | Project coordinator [R3] |
| B | Voice and NLP based tool for diagnostics | Developer [R4] |
| B | Voice and NLP based tool for diagnostics | Developer [R5] |
| B | Voice and NLP based tool for diagnostics | Project manager [R6] |
| C | NLP based tool for indoor navigation | Developer [R7] |
| C | NLP based tool for indoor navigation | Developer [R8] |

Data from the cases were gathered using semi-structured interviews, for which the strategy was prepared according to the guidelines of Galletta (2013). The research framework, described in the preceding section, was utilized to construct the research instrument with which the data was collected. The questions prepared for the semi-structured interviews focused on the components of the framework. The interviews were recorded and the transcripts were analyzed for the empirical study. The transcripts were analyzed using a grounded theory (Strauss and Corbin 1998 and later Heath 2004) inspired approach. Each transcript was first analyzed separately, after which the results of the analysis were compared across cases to find similarities. (Vakkuri et al. 2019a)

*4.2 Empirical Results*

The findings of the empirical study conducted using this framework were summarized into four Primary Empirical Conclusions (PECs). The PECs were communicated as follows:

- PEC1 Responsibility of developers and development is under-discussed
- PEC2 Developers recognize transparency as a goal, but it is not formally pursued
- PEC3 Developers feel accountable for error handling on programming level and have the means to deal with it
- PEC4 While the developers speculate potential socioethical impacts of the resulting system, they do not have means to address them.

These results served to further underline the gap between research and practice in the area. Whereas developers were to some extent aware of some of the goals of the AI ethics principles, these were seldom formally pursued in any fashion. (Vakkuri et al. 2019a)

## 5 Discussion

Rather than discussing the implications of the empirical findings of the study utilizing this framework (as was already done in (Vakkuri et al. 2019a), we discuss the research framework and its implications. As extant studies on AI ethics have been largely conceptual, and e.g. the IEEE EAD guidelines have remarked that much work was still needed to bridge the gap between research and practice in the area (IEEE 2019), this framework provides an initial step towards bridging the gap in this area. It provides a starting point for empirical studies in the area by highlighting important constructs and themes to e.g. discuss in interviews.

However, as this framework was constructed in late 2017, it is now two years old. Since its inception, much has happened in the field of AI ethics and AI in general. The discussion has progressed, and whereas in 2017 the ART model was the current topic of discussion and Fairness an emerging construct, now Fairness has also become a central construct in AI ethics discussion (e.g ACM Conference on Fairness, Accountability, and Transparency, https://fatconference.org/) , especially in the discourse of the United States and the Anglosphere.

Moreover, a recent EU report on Trustworthy AI systems (AI HLEG 2019) discussed *Trustworthiness* as a goal for AI systems, presenting another potentially important construct for the field. However, trustworthiness differs from the existing constructs in that it is not objective and even more difficult to build into a system. Whereas transparency is a tangible attribute of a system or a project that can be evaluated, trustworthiness is ultimately attributed to a system (and its socio-economic context) by an external stakeholder. E.g., a member of the general public may trust or distrust a system, considering it trustworthy.

The discussion on principles in the field continues to be active. Morley et al. (2019) recently proposed a new set of constructs intended to summarize the discussion thus far. Only time will tell whether this novel set of constructs becomes as widely used as the existing constructs such as transparency.

Yet, we maintain that it is pivotal that attempts such as this are made to bring empiricism into this otherwise highly theoretical discussion. Although the field is still evolving, the industry is not waiting for the discussion to finish. AI systems are developed with or without the involvement of AI ethics. To this end, even if the academia does not act, governments and other national and supranational organizations are drafting their own guidelines (E.g. AI HLEG, 2019) and regulations (e.g. https://www.nytimes.com/2019/05/14/us/facial-recognition-ban-san-francisco.html) for AI systems. The academia should aim to participate in this discussion and these actions, even without a unified consensus on the key constructs and principles for AI ethics. As such, the framework presents one way to approach this area of research through an empirical lens.

The framework nonetheless does require further development. Aside from including constructs such as fairness, we argue that it should be *essentialized*. Essentializing refers to a process discussed by Jacobson,

Lawson, Ng, McMahon & Goedicke (2017) in the context of the Essence Theory of Software Engineering (Jacobson, Ng, McMahon, Spence & Lidman 2012) where a Software Engineering (SE) practice is essentialized. Essentialization, according to Jacobson et al. (2017), refers to the process of distilling e.g. a software engineering practice into its essential components in order to communicate it clearly and in a unified fashion, while communicating it in the Essence language.

In the context of this framework, we see essentialization as one way to make it more understandable for industry experts. Essentializing a practice, or method, or a framework has three steps, according to Jacobson et al. (2017):

1. *Identifying the elements - this is primarily identifying a list of elements that make up a practice. The output is essentially a diagram.*
2. *Drafting the relationships between the elements and the outline of each element - at this point, the cards are created.*
3. *Providing further details - Usually, the cards will be supplemented with additional gui-delines, hints and tips, examples, and references to other resources, such as articles and books*

In this fashion, the framework could be essentialized by e.g. making Essence *alphas* out of the principles such as transparency. Alphas, in the context of Essence, are things to work with which are measured in order to see progress on the endeavor (Jacobson et al. 2012). One could thus consider them goals. The framework could then be extended by practices which seek to help an organization progress in achieving these ethical principles.

As it stands, the framework can be utilized for empirical studies in the area of AI ethics. It presents a practice-focused view of AI ethics. However, it does not cover all the aspects of the AI ethics discussion in 2019 (and beyond). Depending on the context, one may wish to extend it to include fairness as the fourth key principle for AI ethics, and/or trustworthiness.

# 6 Conclusions & Future work

In this paper, we have presented a framework for approaching AI ethics through practice. Having conducted an empirical study using the framework (Vakkuri et al. 2019a), we discussed the implications of the framework and how it should be developed further in this paper. Though the framework, as is, can be utilized for empirical studies, it should be complemented by the inclusion of some of the more recent AI ethics constructs such as fairness and trustworthiness to make it more current. Given that the framework was originally devised in late 2017, the discussion in the field of AI ethics has since then gone forward.

We seek to develop the framework further ourselves. We have utilized a similar framework in another study (Vakkuri et al. 2019b). Aside from simply expanding the framework to include fairness and trustworthiness, we have plans to *essentialize* the framework by utilizing the Essence Theory of Software Engineering (Jacobson et al. 2012, Jacobson et al. 2017) in order to make it more relevant to practitioners.